\documentclass{iau} 
\usepackage{graphicx}

\title[PSR B0943+10] 
{X-rays from the mode-switching  PSR~B0943+10}

{

\author[Sandro Mereghetti, et al.]{S. Mereghetti$^1$, L. Kuiper$^2$, A. Tiengo$^{3,1,4}$, J.  Hessels$^{5,9}$, W.~Hermsen$^{2,9}$,   K. Stovall$^6$, A. Possenti$^7$, J. Rankin$^8$,   P. Esposito$^9$, R.~Turolla$^{10,11}$,  D. Mitra$^{8,12,13}$, G. Wright$^{14}$, B.~Stappers$^{14}$, A.~Horneffer$^{15}$,  S. Oslowski$^{15,16,17}$, M. Serylak$^{18,19}$, J.-M.~Griessmeier$^{20,19}$, M. Rigoselli$^{1,21}$}

\affiliation{$^1$ INAF-IASF Milano, Italy \\
$^2$ SRON,    Utrecht, The Netherlands \\
$^3$ Scuola Universitaria Superiore IUSS Pavia,    Italy \\
$^4$ INFN, Sezione di Pavia,  Italy \\
$^5$ ASTRON,  Dwingeloo, The Netherlands \\
$^6$ Department of Physics and Astronomy, University of New Mexico, Albuquerque, NM, USA \\
$^7$ INAF - Osservatorio Astronomico di Cagliari,   Selargius, Italy  \\
$^8$ Physics Department, University of Vermont, Burlington, VT 05405, USA \\
$^9$ Anton Pannekoek Institute for Astronomy, Univ. of Amsterdam, The Netherlands \\
$^{10}$ Dipart. di Fisica e Astronomia, Universit\`a di Padova, Italy \\
$^{11}$ MSSL-UCL, Holmbury St. Mary, Dorking, UK \\
$^{12}$ National Centre for Radio Astrophysics, Ganeshkhind, Pune,  India \\
$^{13}$ Janusz Gil Institute of Astronomy, Univ. of Zielona G\'ora,   Poland \\
$^{14}$ Jodrell Bank Centre for Astrophysics,   Univ. of Manchester,   UK \\
$^{15}$ Max-Planck-Institut f{\"u}r Radioastronomie, Bonn, Germany \\
$^{16}$ Fakult{\"a}t f{\"u}r Physik, Universit{\"a}t Bielefeld, Germany \\
$^{17}$ currently at    Swinburne Univ. of Technology, Australia \\
$^{18}$ Dept. of Physics \& Astronomy, Univ. of the Western Cape,  Bellville, South Africa \\
$^{19}$ Station de Radioastronomie de Nan\c{c}ay, Observatoire de Paris,  CNRS,   Nan\c{c}ay, France \\
$^{20}$  LPC2E - Universit\'{e} d'Orl\'{e}ans, France \\
$^{21}$ Universit\`a di Milano Bicocca, Milano, Italy
}

\pubyear{2017}
\volume{337}  
\setcounter{page}{1}
\jname{Pulsar Astrophysics -­ The Next 50 Years}
\editors{P. Weltevrede, B.B.P. Perera, L. Levin Preston \& S. Sanidas, eds.}

\def \xmm {\emph{XMM-Newton} }

\newcommand{\bc}{\begin{center}}
\newcommand{\ec}{\end{center}}
\def\pdot {\dot P}

\def\ltsima{$\; \buildrel < \over \sim \;$}
\def\lsim{\lower.5ex\hbox{\ltsima}}
\def\gtsima{$\; \buildrel > \over \sim \;$}
\def\gsim{\lower.5ex\hbox{\gtsima}}

\def\src {PSR\,B0943+10}
\def\psr {PSR\,B0943+10}

\begin{document}

\maketitle

\begin{abstract}
New simultaneous X-ray and radio observations of the archetypal mode-switching  pulsar \psr\ have been carried out with  {\it XMM-Newton}  and the LOFAR, LWA and Arecibo radio telescopes in November 2014. 
They allowed us to better constrain the  X-ray spectral and variability properties of this pulsar and to detect, for the first time, the  X-ray pulsations also during the  X-ray-fainter mode. 
The combined timing and spectral analysis indicates that unpulsed non-thermal emission, likely of magnetospheric origin,   and pulsed thermal emission from a small polar cap are present during both radio modes and  vary  in a correlated way.

\end{abstract}

\firstsection 
\section{Introduction}

Until recently, the study of variability in rotation-powered pulsars was limited to the radio band,
where changes in the emission properties related to phenomena occurring  on very different time scales, from sub-pulses to long-term moding and intermittence,  are routinely observed.
On the contrary, rotation-powered neutron stars were generally  believed to have a steady  luminosity  at X- and $\gamma$-ray energies,
to the point that they are typically used as calibration sources.
Some X-ray variability, in the form of short bursts associated with  X-ray  flux enhancements and spectral changes, has been observed in two allegedly rotation-powered pulsars (\cite[Gavriil \etal\ 2008; Archibald \etal\ 2016]{gav08,arc16}), and  some long term X-ray variations have also been observed in one member of the XDIN class (\cite{hol12}). However, all these variable phenomena are believed to result from magnetic activity and are not expected to occur in old radio pulsars with ``canonical'' magnetic fields (\cite[see, e.g., Mereghetti, Pons \& Melatos 2015]{mer15}).

The situation has changed with the recent discovery of  variations in the X-ray flux of  \src\ 
(\cite[Hermsen \etal\ 2013]{her13}), one of the mode-switching pulsars best studied in the radio band.
Here, we briefly review the most important results obtained for \psr\  and compare its properties with those of other mode-switching pulsars recently observed at X-ray energies.

\begin{figure}[bh]
\begin{center}
 \includegraphics[width=10cm]{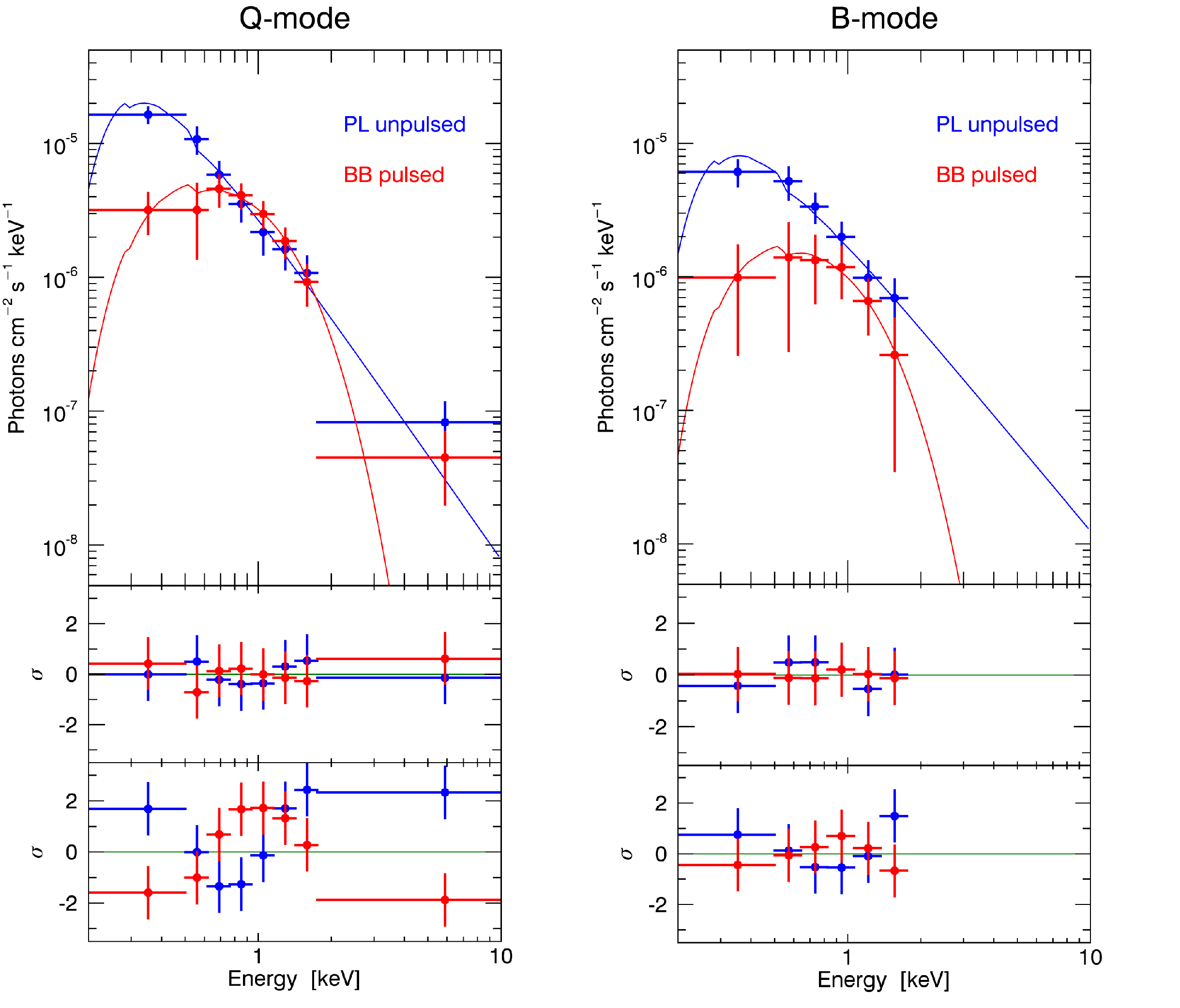} 
 \caption{X-ray spectra of the unpulsed (blue) and pulsed (red) emission from  \psr\ in the Q-mode (left) and B-mode (right). In the top panels  the unpulsed flux is fitted with a power law and the pulsed flux with a blackbody. The corresponding residuals are shown in the middle panels.  The bottom panels, which show the residuals obtained by fitting the unpulsed flux with a blackbody and the pulsed flux with a power law, demonstrate that this fit is acceptable for the B-mode but not for the Q-mode. 
}
\label{fig_spectra}
\end{center}
\end{figure}

\section{X-ray results}
       
\psr\  ($P$\,=\,1.1 s,  $\pdot$\,=\,3.5$\times$10$^{-15}$ s s$^{-1}$) is a relatively old pulsar ($\tau$\,=\,5 Myr)  with a  dipole magnetic field of 4$\times$10$^{12}$ G. 
In the radio band, at irregular intervals of $\lsim$ few hours, it alternates  between a radio bright (B-) mode, characterized by regularly drifting subpulses,  and a  quiescent (Q-) mode in which the radio emission is on average fainter and the subpulses have a chaotic pattern (\cite[Suleimanova \& Izvekova 1984; Bilous 2017]{sul84,bil17}).

The faint X-ray emission from \psr\ (few 10$^{-15}$ erg\,cm$^{-2}$\, s$^{-1}$) was discovered in 2003 (\cite[Zhang \etal\ 2005]{zha05}), but it could be studied in detail only when longer observations with simultaneous radio coverage were obtained in 2011. This made it possible to analyse separately the X-ray data of the two radio modes, leading to the discovery of X-ray variability anti-correlated with the radio flux (\cite[Hermsen \etal\ 2013]{her13}). 
It was found that, during the {\it radio-fainter} Q-mode,  the X-ray flux is brighter by a factor $\sim$2.5, shows X-ray pulsations, and has a blackbody plus power-law spectrum.   
In the B-mode, the spectrum was less constrained, being fit equally well by either a blackbody or a power law.

We performed a new observing campaign in November 2014, using \xmm and the LOFAR, LWA and Arecibo radiotelescopes (\cite[Mereghetti \etal\ 2016]{mer16}),  
obtaining X-ray esposures ($\sim$120 ks in Q- and $\sim$175 ks  B-mode) longer by factors of 2.4 and 3.5 than those of the previous data, respectively.
This allowed us to better constrain the spectral properties of \psr\ and to detect its  X-ray  pulsations also during the  B-mode.
The new data show that the phase-averaged spectrum is well fit by the sum  of a power law with photon index $\sim$2.4 and a blackbody in both radio modes. The blackbody temperatures in the two modes are $kT_Q$\,=\,0.27$\pm$0.04  keV and  $kT_B$\,=\,0.24$\pm$0.03   keV and the emitting radii are of the order of $\sim$30 m  (for  d\,=\,890 pc).

By applying a maximum likelihood spectral analysis which used also the timing information of the detected counts, we   derived the spectra of the pulsed and unpulsed emission.  
As shown in Fig.~\ref{fig_spectra} (left), we confirmed that in the  Q-mode  the pulsed flux is thermal while the unpulsed flux is non-thermal.
The blackbody describing  the pulsed emission and the power law describing the unpulsed emission are consistent with the two components used to fit the total spectrum of the Q-mode.

\begin{table}
  \begin{center}
  \caption{Comparison of three mode-switching pulsars observed in X-rays}
  \label{tab1}
 {\scriptsize
  \begin{tabular}{|l|c|c|c|}\hline 
     & {\bf \psr\ } & {\bf PSR\,B0823+26} & {\bf PSR\,B1822--09 }  \\ 
   \hline

$P$  [s]                      &  1.1    & 0.53    & 0.77     \\ 
$\pdot$ [s s$^{-1}$]  &   $3.5\times10^{-15}$  &   $1.7\times10^{-15}$    &   $5.3\times10^{-14}$    \\ 
$\tau$  [Myr]             & 5    &  5    &  0.2 \\
$B$  [G]                    &  $2\times10^{12}$ & $9.6\times10^{11}$ &  $6.5\times10^{12}$ \\  
$\dot{E}_{rot}$ [erg s$^{-1}$]  &  10$^{32}$   &  $4.5\times10^{32}$   &  $4.6\times10^{33}$      \\
\hline
    Distance  [kpc]      &  0.89  & 0.31   & 0.26  \\  \hline
Likely geometry  &    $\sim$aligned  & $\sim$orthogonal  &   $\sim$orthogonal \\ \hline
Mode duration  &   $\sim$hours &  $\sim$hours (also nulling)  &   few minutes \\\hline
Radio interpulse           &  No  &  Yes  &  Yes (in Q-mode) \\\hline
 Radio pre/postcursor      &  precursor in Q-mode  & postcursor in B-mode & in B-mode\\  \hline
 Drifting subpulses               &   in B-mode &  Yes  &  No  \\\hline
 X-ray  flux       &  brighter in Q-mode   &  brighter in B-mode  &  constant\\ \hline
$L_X$ [erg s$^{-1}$]  &  $\sim2\times10^{29}$    &  $\sim2\times10^{28}$  &  $\sim3\times10^{29}$      \\ \hline 
Total X-ray spectrum & BB+PL (or BB+BB) & BB+BB (or BB+PL)  & BB+BB  \\ \hline
Spectral-timing  & BB pulsed     &  favors BB+BB &   both BB pulsed \\
analysis             &  PL unpulsed &                       &                            \\ \hline
References   & Mereghetti \etal\ 2016 & Hermsen \etal\ 2018  & Hermsen \etal\ 2017  \\ \hline

                 \end{tabular}
  }
 \end{center}
\end{table}

The situation is less well constrained for the B-mode:  for both the pulsed and unpulsed emission, it is impossible to distinguish between a  power law and a blackbody, since they give similarly good fits (Fig.\,\ref{fig_spectra}, right).
However,  it is reasonable to assume that the same two components seen in the Q-mode are present, although with a reduced flux, also during the B-mode. In other words, we attribute the  pulsations to the thermal component in both modes.  When the pulsar changes from the Q to the  B-mode, the  thermal flux decreases by a factor $\sim$3 and the non-thermal one decreases by a factor $\sim$2.

\section{Discussion}

The results obtained in the  2014 campaign  indicate that the X-ray properties of \psr\ in the two radio modes are not as different as it was   previously thought: a blackbody plus power law spectrum and X-ray pulsations  (which can be attributed to the thermal component) are present in both modes, although with different fluxes.  
Thus, it is not necessary to invoke global magnetospheric changes capable of entirely suppressing one of the spectral components (\cite[Hermsen \etal\ 2013; Mereghetti \etal\ 2013]{her13,mer13}) when the pulsar switches to its X-ray fainter mode.
The small emitting area inferred from the blackbody fits is broadly consistent with the polar cap size expected in the partially screened gap model,  which is smaller than that of a pure dipole owing to multipolar field components (\cite[Gil \etal\ 2007]{gil07}). 

The mechanism causing the mode-switching in radio pulsars is still unknown and  the study of X-ray variability might help to understand the origin of  this phenomenon. 
Indeed, the interesting results obtained for \psr\ prompted X-ray observations of other mode-switching pulsars. 
Unfortunately, these old and not very energetic pulsars are faint in  X-rays. Up to now, \xmm data have been obtained for PSR\,B1822--09 (\cite[Hermsen \etal\ 2017]{her17}) and PSR\,B0823+26 (\cite[Hermsen \etal\ 2018]{her18}).
Their properties  are compared with those of \psr\ in Table\,1. 
 
PSR B1822--09, younger  and more energetic than \psr ,  shows a main radio pulse and a radio interpulse separated by $\sim0.5$ in phase and, contrary to \psr ,  is believed to be a nearly orthogonal rotator.
Its  Q- and a B-mode radio states last only a few minutes and  no evidence for X-ray variations was found.
Its X-ray spectrum is fit by the sum of two blackbodies with $kT_1$\,=\,0.08 keV,  $kT_2$\,=\,0.19 keV  and emitting radii  $R_1$ = 2 km, $R_2$ = 100 m.  
Both the hotter and the cooler  blackbody components seem to contribute to the X-ray pulsations, which are present in both radio modes.
The  lack of X-ray variability could be due to  the short duration of the radio modes. This could arise, e.g., if the timescale for the X-ray variation is longer than the mode duration.
The timing parameters of PSR B0823+26 are very similar to those of \psr . It also  shows a bright and a quiet radio mode (\cite[Sobey \etal\ 2015]{sob15}), but, contrary to \psr , the X-ray flux is higher during the B-mode.

Altough the global picture emerging from these data is still puzzling, the heterogeneous pattern of X-ray/radio variations seen in these few objects, anticipates  that future, more sensitive observations will reveal a rich phenomenology that can potentially provide interesting clues on the mode-switching mechanism, and more in general, on the emission processes of rotation-powered pulsars.

\end{document}